\documentclass[%
 aps,
 amsmath,amssymb,
 aps,
]{revtex4}
\usepackage{epsfig, color}

\usepackage{graphicx}
\graphicspath{%
    {converted_graphics/}
    {/}
}
\usepackage{amsfonts,amssymb,
amsthm}
\usepackage{amsmath}
\usepackage{graphicx}
\usepackage{enumerate}
\usepackage{subfigure}
\usepackage{mathtools}
\usepackage{amscd}
\usepackage[toc]{appendix}
\usepackage{afterpage}
\usepackage{bm}
\usepackage{epstopdf}
\usepackage{mathrsfs}
\newcommand{\eproof}{\hfill{\vrule height5pt width5pt depth0pt}\medskip}

\renewcommand{\proof}{{\noindent {\em Proof}.\;}}

\newtheorem{theorem}{Theorem}[section]

\newtheorem{definition}[theorem]{Definition}

\begin{document}

\title[Lane-Emden equation]
{Stability analysis  of  orbital modes  for a
generalized Lane-Emden equation}

\author{Ronald Adams}
\email{adamsr25@erau.edu}
\affiliation{Department of Mathematics, Embry-Riddle Aeronautical University, Daytona Beach, FL 32114-3900, USA}

\author{Stefan C. Mancas}
\email{mancass@erau.edu}
\affiliation{Department of Mathematics, Embry-Riddle Aeronautical University, Daytona Beach, FL 32114-3900, USA}

\author{Haret C. Rosu}
\email{hcr@ipicyt.edu.mx}
\affiliation{IPICYT, Instituto Potosino de Investigacion Cientifica y Tecnologica,\\
Camino a la presa San Jos\'e 2055, Col. Lomas 4a Secci\'on, 78216 San Luis Potos\'{\i}, S.L.P., Mexico}

\begin{abstract}
We present a stability analysis of the standard nonautonomous systems type for a recently introduced generalized Lane-Emden equation which is shown to explain the presence of some of the structures observed in the atomic spatial distributions of magnetically-trapped ultracold atomic clouds. A Lyapunov function is defined which helps us to prove that stable spatial structures in the atomic clouds exist only for the adiabatic index  $\gamma=1+1/n$ with even $n$.  In the case when $n$ is odd we provide an instability result indicating the divergence of the density function for the atoms. Several numerical solutions, which according to our stability analysis are stable, are also presented.

\end{abstract}
\maketitle

\section{The generalized Lane-Emden equation}

In a recent article \cite{Rod1},  Rodrigues {\it et al.} use the condition of hydrostatic equilibrium in the set of the fluid continuity and momentum (Navier-Stokes) equations and a Poisson-like equation to determine the equation of state of a laser-cooled gas within a magneto-optical trap. The authors obtain what they call  the {\em generalized}
Lane-Emden equation for the confined atomic profiles, which is  also derived in the context of astrophysical fluids \cite{Har}, and  is given by
\begin{equation}\label{eq1}
\gamma \frac{1}{\zeta^2}\frac{d}{d \zeta}\left(\zeta^2 \theta^{\gamma-2}\frac {d\theta}{d\zeta} \right)+1-\Omega \theta=0.
\end{equation}
The   adiabatic index  $\gamma$  relates the pressure and density of atoms by the relation $p=C_\gamma \rho^{\gamma}$,  where $C_\gamma=\frac{p(0)}{\rho(0)^\gamma}$, and  $p(0)$,  $\rho(0)$ are the pressure and density  at the center of  the atomic cloud while the  density at some distance $r$ from the center is given by   $\rho(r)=\rho(0)\theta(r)$.  Using radial symmetry,  $\theta(\zeta)$ is the nondimensional density of atoms which depends on the nondimensional  distance $\zeta$ from the center of the cloud given by $\zeta=\frac{r}{a_\gamma}$, with $a_\gamma=\sqrt{\frac{C_\gamma}{3 m \omega_0^2}}\rho(0)^{\frac{\gamma-1}{2}}$, and trap frequency  $\omega_0^2 =\frac{\kappa}{m}$ \cite{Rod1}. The nondimensional constant $\Omega$   is the ratio of multiple scattering induced radiation pressure to trapping forces,  is  defined by $\Omega=\frac{Q\rho(0)}{3 m \omega_0^2}$, with $Q$ being the square of effective electric charge  \cite{Pruv},  and is the equivalent of the plasma frequency \cite{Men}. The stability of solutions depends on $\Omega$, as we will see in the following section, with stable solutions found  when $0<\Omega<1$.
Ter\c cas {\it et al.}   \cite{Terc}  derived the same  equation which models  the polytropic equilibrium of a magneto-optical trap that  describes the crossover
between the two limiting cases: temperature-dominated ($\Omega \rightarrow 0$), and multiple-scattering-dominated traps  ($\Omega \rightarrow 1$).

For  $\gamma=2$,  (\ref{eq1}) is  linear
\begin{equation}\label{op}
\theta_{\zeta\zeta}+\frac 2 \zeta \theta_\zeta+\frac{1-\Omega \theta}{2}=0,
\end{equation}
and  may be roughly considered as a repulsive electrostatic counterpart of the classic Lane-Emden equation in astrophysics with solutions representing the Newton-Poisson gravitational potential of stars, considered as spheres filled with polytropic gas \cite{Emden}.
For other values of $\gamma$, (\ref{eq1})  has an additional operatorial term of the form $\gamma(\gamma-2){\theta_\zeta}^2/\theta$, and also two nonoperatorial terms instead of one.

When the effects of the multiple scattering can be neglected, and thermal effects dominate $\Omega \rightarrow 0$, and for $\gamma \ne1$  the atomic density is  given by
\begin{equation}\label{eq1a}
\theta(\zeta)=\left[(1-\gamma)\left(c_1+\frac{c_2}{\zeta}+\frac{\zeta^2}{6\gamma}\right)\right]^{\frac{1}{\gamma-1}}.
\end{equation}
For a bounded density at the origin $c_2=0$, and using $\theta(0)=\theta_0>0$  the solution to  \eqref{eq1a}   corresponding  to the density of atoms having no effective charge is
\begin{equation}\label{eq1b}
\theta(\zeta)=\left[{\theta_0}^{\gamma-1}-\frac{\gamma-1}{6\gamma}\zeta^2\right]^{\frac{1}{\gamma-1}}.
\end{equation}
The case $\gamma=1$ corresponds to an  isothermal gas as discussed in \cite{Rod1}, and yields  the Maxwell-Boltzmann equilibrium where (\ref{eq1}) reduces to the separable equation
\begin{equation}\label{eq1c}
\frac{d}{d \zeta}\left(\zeta^2 \frac{d \ln \theta}{d \zeta} \right)=-\zeta^2.
\end{equation}
 By two quadratures, and assuming the same initial  conditions, the atomic density has the Gaussian profile
\begin{equation}\label{eq1d}
\theta(\zeta)=\theta_0e^{-\frac{\zeta^2}{6}}.
\end{equation}
This solution can also be found by taking the limit $\gamma \rightarrow 1 $ in (\ref{eq1b}), and describes a trap in the temperature-limited regime when  scattering effects are negligible \cite{Rod2}.

On the other hand, when the effects of multiple scattering dominate then $\Omega\rightarrow 1$.  For an isobaric process, $\gamma=0$,  the density of atoms is given by
\begin{equation}
 \theta(\zeta)=\frac 1 \Omega H(\xi-\xi_0),
\end{equation}
where $H(\xi-\xi_0)$ is the Heaviside step  function, and $\xi_0=\frac{3}{\sqrt[3]{4 \pi \Omega}}$ is the Lane-Emden radius of the traps, which was also observed in the  experiments of \cite{Gatto}. This solution is known as the {\em water-bag}  equilibrium profile \cite{Wal,Rod2}.

For the aforementioned case $\gamma=2$, which can be taken into account in the astrophysics of
dark matter halos with possible substantial pressure compared to the atom density of the {\em dark} atoms \cite{Bhar,Har}, the polytropic equation of state is $p=C_2\rho^2$.  Then (\ref{op}) can be written in self-adjoint form
\begin{equation}\label{eq2b}
\frac{d}{d\zeta}\left(\zeta^2\frac{d\theta}{d\zeta}\right)+\zeta^2\frac{1-\Omega \theta}{2}=0~
\end{equation}
and if we use $\psi=\frac{1-\Omega\theta}{2}$, and $A=-\frac \Omega 2$, it becomes
\begin{equation}
 \frac{d}{d\zeta}\left(\zeta^2\frac{d\psi}{d\zeta}\right)+A\zeta^2\psi=0.
 \end{equation} 
In the latter format, this equation is close to the `electrostatic Lane-Emden equation', known as the Thomas-Fermi equation,  associated with the field of the statistical distribution of electrons in heavy atoms which takes into account their repulsive electrostatic interactions \cite{Thomas}
 \begin{equation}
 \frac{d}{d\zeta}\left(\zeta^2\frac{d\psi}{d\zeta}\right)+B\zeta^2\psi^{\frac{3}{2}}=0~,
 \end{equation}
 where $B=-8\sqrt{2}/3\pi$, see equation (1.2) in \cite{Thomas}.

By assuming  initial conditions $\theta(0)=\theta_0$, and $\theta_\zeta(0)=0$,  the atom density is given by
\begin{equation}\label{e3}
\theta(\zeta)=\frac{1}{\Omega}\Bigg[1+(\theta_0\Omega-1)\,{\rm shc}\left(\sqrt{\frac{\Omega}{2}}\zeta\right)
\Bigg]~,
\end{equation}
where
${\rm shc}$ denotes the cardinal hyperbolic sinus function defined by
$$
{\rm shc}(x):=  \left\{
\begin{array}{ll}
\frac{\sinh (x)}{x}, & \ \  \mbox{for}\  x\ne 0, \\
1,& \ \  \mbox{for}\  x=0.
\end{array}
 \right.
$$
The boundary of the halo $\zeta_M$, can be found numerically by assuming that on the boundary we must have zero atom density, thus
${\rm shc}\left(\sqrt{\frac{\Omega}{2}}\zeta_M\right)=\frac{1}{1-\theta_0\Omega}$,
provided that $0<\Omega< \frac{1}{\theta_0}$.

Since  \eqref{eq1} cannot be solved analytically for other values of $\gamma$, an important issue that we address in the rest of this paper is the stability analysis of solutions that are obtained numerically. We develop this stability analysis using a generalized Lyapunov function for the corresponding nonautonomous system of differential equations in Section 2. If the positive departure from unity of the adiabatic index is parametrized by $1/n $, where $n\in \mathbb{N}$, then we find that stable solutions exist only for even $n$, while for odd $n$ all the solutions are unstable. The numerical solutions are presented in Section 3 as a proof of the results obtained in the previous section, and the paper ends  with a conclusion section.

\section{The stability analysis}

For any $\gamma \ne 1$, we write (\ref{eq1}) in self-adjoint form as
 \begin{equation}\label{eq2}
\frac{d}{d \zeta}\left(\zeta^2 \frac{d}{d \zeta}\theta ^{\gamma -1}\right)=\frac{\gamma-1}{\gamma}\zeta^2\left (\Omega \theta-1\right)~.
\end{equation}
Letting $\theta=z^{\frac{1}{\gamma -1}}$ we obtain
\begin{equation}\label{eq3}
\frac{d}{d \zeta}\left(\zeta^2 \frac{dz}{d\zeta}\right)=\frac{\gamma-1}{\gamma}\zeta^2\left (\Omega z^{\frac{1}{\gamma-1}}-1\right),
\end{equation}
and by using  $\gamma=1+1/n$ with $n\in \mathbb{N}$, we get
\begin{equation}\label{eq4}
\frac{d}{d \zeta}\left(\zeta^2 \frac{dz}{d\zeta}\right)=\frac{\zeta^2}{n+1}\left (\Omega z^n-1\right).
\end{equation}
Eq. \eqref{eq4} can be written as a nonautonomous first order system
\begin{equation}\label{eq5}
\frac{dy}{d\zeta}=f(\zeta,y)
\end{equation}
using
$$
y=\begin{pmatrix}
z\\
z_\zeta\\
\end{pmatrix}, \quad  {\rm and} \quad
f(z,z_\zeta)=\begin{pmatrix}
z_\zeta\\
\frac{1}{n+1}(\Omega z^n-1)-\frac{2z_\zeta}{\zeta}
\end{pmatrix}.
$$
There is only one real critical point of the system (\ref{eq5}) for  $n$ odd given by ${\bar{y}_0}^T=\left(\frac{1}{\Omega^{1/n}},0\right)$
while  for  $n$  even there are two ${\bar{y}_{1,2}}^T=\left(\mp\frac{1}{\Omega^{1/n}},0\right)$.
All the critical points are shifted  to the origin by the substitution $x^{T}=y^{T}-{\bar{y}}^{T}=\left(x_1,x_2\right)$, then \eqref{eq5} becomes
\begin{equation}\label{system1}
x_\zeta=f(\zeta,x+\bar{y})=\begin{pmatrix} x_2\\ \frac{1}{n+1}\left[\Omega\left(x_1\mp\frac{1}{\Omega^{1/n}} 
\right)^n-1\right]-\frac{2x_2}{\zeta}\\\end{pmatrix}.
\end{equation}
First, we define the notion of stability for a general nonlinear nonautonomous system
\begin{align}\label{system2}
x_\zeta=f(\zeta,x),\quad \zeta\geq \zeta_0,
\end{align}
where $\zeta_0\geq0$ and $x(\zeta_0)=x_0\in\mathbb{R}^n$.  We assume $f:\mathbb{R}^{+}\times\mathbb{R}^n\rightarrow\mathbb{R}^{n}$ is continuous and satisfies $f(\zeta,0)=0$, $\forall \zeta\geq \zeta_0$.
\begin{definition}\em{
(\cite{HK}) The system (\ref{system2}) is said to be:
\begin{enumerate}
\item[(i)] \textit{Lyapunov stable} if for each $\epsilon>0,$ there is $\delta(\epsilon,\zeta_0)>0$ such that
\begin{align}\label{LI}
\left\|x(\zeta_0)\right\|<\delta\Rightarrow \left\|x(\zeta)\right\|<\epsilon,\quad \forall \zeta\geq \zeta_0\geq 0.
\end{align}
\item[(ii)] \textit{Uniformly stable} if for each $\epsilon>0$, there is $\delta=\delta(\epsilon)>0$, independent of $\zeta_0$, such that (\ref{LI}) is satisfied.
\item[(iii)]\textit{Asymptotically stable} if it is \textit{Lyapunov stable},  and there is a constant $l=l(\zeta_0)>0$ such that $x(\zeta)\rightarrow 0$ as $\zeta\rightarrow\infty$, for all $\left\|x(\zeta_0)\right\|<l$.
\item[(iv)] \textit{Uniformly asymptotically stable} if it is \textit{uniformly stable}, and there is a constant $l>0$ independent of $\zeta_0$, such that for all $\left\|x(\zeta_0)\right\|<l$, \; $x(\zeta)\rightarrow 0$ as $\zeta\rightarrow\infty$, uniformly in $\zeta_0$.  This means that for each $\eta>0$ there is $T=T(\eta)>0$ such that
\begin{align}
\left\|x(\zeta)\right\|<\eta,\quad\forall \zeta\geq \zeta_0+T(\eta),\;\forall\left\|x(\zeta_0)\right\|<l.
\end{align}
\item[(v)]\textit{Exponentially stable} if there exist positive constants $l$, $k$, and $p$ such that
\begin{align}
\left\|x(\zeta)\right\|\leq k\left\|x(\zeta_0)\right\|e^{-p(\zeta-\zeta_0)},\quad \forall \left\|x(\zeta_0)\right\|<l.
\end{align}
\end{enumerate}}
\end{definition}
Before attempting the stability problem for the nonlinear system \eqref{system1}, it behooves us to first study the stability of the linearization of \eqref{system1} about $\bar{x}^{T}=\left(0,0\right)$.  Supposing $n$ is even,  we use the left equilibrium ${\bar{y}_1}^{T}=\left(- \frac{1}{\Omega^{1/n}},0\right)$, and we set $A(\zeta)=\nabla_{x} f(\zeta,x)\mid_{x=\bar{x}},$  then
\begin{align}\label{A}
\left.A(\zeta)=\begin{pmatrix} 0 & 1\\ \frac{\Omega}{1+1/n}\left(x_1-\frac{1}{\Omega^{1/n}}\right)^{n-1} & 
 -\frac{2}{\zeta}\\\end{pmatrix}\right|_{x=\bar{x}}=\begin{pmatrix} 0 & 1\\-\frac{1}{1+1/n}\Omega^{1/n} & -\frac{2}{\zeta}\\\end{pmatrix}.
\end{align}
Using  \eqref{A} we then consider the corresponding linear distance-varying system to \eqref{system1},
\begin{align}\label{tvls}
x_\zeta=A(\zeta)x(\zeta),\quad x(\zeta_0)=x_0.
\end{align}
To establish the stability of \eqref{tvls}, we make use of a linear matrix inequality developed in \cite{CY}.  We restate the result below for completeness.
\begin{theorem}\em{(Theorem~1 in \cite{CY}.)
Consider the system (\ref{tvls}).  Suppose there exists a positive definite differentiable matrix function $P:\mathbb{R}^{+}\rightarrow\mathbb{R}^{n\times n}$ and a continuous function $g:\mathbb{R}^{+}\rightarrow\mathbb{R}$ such that
\begin{align}\label{hyp1}
A^{T}P+PA+P_\zeta\leq g(\zeta)P
\end{align}
with $\int_{\zeta_0}^{\infty}g(\zeta)\;d\zeta=-\infty$, then the system (\ref{tvls}) is asymptotically stable.}
\end{theorem}
\begin{theorem}\em{
The origin is asymptotically stable for the system \eqref{tvls}.}
\end{theorem}
\proof
To apply the above result to the linear system~\eqref{tvls} let $P(\zeta)=\bigl(\begin{smallmatrix} a(\zeta) & b(\zeta)\\ b(\zeta) & c(\zeta)\end{smallmatrix} \bigr)$, then
\begin{align}\label{Q}
A^{T}P+PA+P_\zeta=\begin{pmatrix}a_{\zeta} -\frac{\Omega^{1/n}}{1+1/n}b& a-\frac{2}{\zeta}b+b_{\zeta}-\frac{\Omega^{1/n}}{1+1/n}c\\ a-\frac{2}{\zeta}b+b_{\zeta}-\frac{\Omega^{1/n}}{1+1/n}c&
2b-\frac{4}{\zeta}c+c_\zeta
\end{pmatrix}.
\end{align}
By setting $b=0$, $a=\frac{1}{\zeta}$, $c=\frac{1+1/n}{\Omega^{1/n}\zeta}$, and $g(\zeta)=-\frac{1}{\zeta}$ the matrix inequality (\ref{hyp1}) becomes
\begin{align}\label{hyp2}
\begin{pmatrix}-\frac{1}{\zeta^2} & 0\\0 & -\frac{5(1+1/n)}{\Omega^{1/n}\zeta^2}
\end{pmatrix}\leq\begin{pmatrix} -\frac{1}{\zeta^2} & 0\\0 & -\frac{1+1/n}{\Omega^{1/n}\zeta^2}\end{pmatrix},
\end{align}
hence  for $\zeta_0>0$,  (\ref{hyp2}) is satisfied for $\zeta\geq\zeta_0$. Furthermore, the matrix function
\begin{align*}
P(\zeta)=\begin{pmatrix} \frac{1}{\zeta} & 0\\0 & \frac{1+1/n}{\Omega^{1/n}\zeta} \end{pmatrix},
\end{align*}
is differentiable. For the positive definiteness of $P(\zeta)$ we compute $x^{T}Px$,
\begin{align*}
x^{T}Px=\begin{pmatrix}x_1 \\ x_2\end{pmatrix}^{T}\begin{pmatrix} \frac{1}{\zeta} & 0\\0 & \frac{1+1/n}{\Omega^{1/n}\zeta} \end{pmatrix}\begin{pmatrix} x_1\\x_2\end{pmatrix}=\frac{1}{\zeta}\left({x_1}^2+\frac{1+1/n}{\Omega^{1/n}}{x_2}^2\right)>0,
\end{align*}
 for $x_1$ and $x_2$ both nonzero.  Therefore the distance-varying system \eqref{tvls} is locally asymptotically stable. \eproof

It is only of interest to us to study the left equilibrium point
${\bar{y}_{1}}^{T}=\left(-\frac{1}{\Omega^{1/n}},0\right)$
for stability, as the equilibrium point ${\bar{y}_{2}}^T=\left(\frac{1}{\Omega^{1/n}},0\right)$ will  be shown to be unstable (see Theorem~\ref{unstable}),  for $n \in \mathbb{N}$.  Before we state the stability/instability results for the nonlinear system \eqref{system1} we first need to introduce some terminology that is used in \cite{SL}.
\begin{definition}\em{(\cite{SL})
A scalar continuous function $V(x)$ is said to be \textit{locally positive definite} if $V(0)=0$ and, in a ball $B_r(0)$
\begin{align}
x\neq0\implies V(x)>0.
\end{align}
If $V(0)=0$ and the above property holds over the whole state space, then $V(x)$ is said to be \textit{globally positive definite}.}
\end{definition}
Related concepts can be defined analogously, in a local or global sense.  A function $V(x)$ is \textit{negative definite} if $-V(x)$ is positive definite;  $V(x)$ is positive \textit{semi-definite} if $V(0)=0$ and $V(x)\geq0$ for $x\neq0$; $V(x)$ is \textit{negative semi-definite} if $-V(x)$ is positive semi-definite.  For our purposes we will need the notion of positive definiteness for a scalar valued distance-varying function.
\begin{definition}\em{(\cite{SL})}
A scalar distance-varying function $V(x,\zeta)$ is \textit{locally positive definite} if $V(0,\zeta)=0$, and there exists a distance-invariant positive definite function $V_0(x)$ such that
\begin{align}
\forall \zeta\geq\zeta_0,\quad V(x,\zeta)\geq V_0(x).
\end{align}
Thus, a distance-variant function is locally positive definite if it dominates a distance-invariant locally positive function.  The notions of negative definite, semi-definite, and negative semi-definite can be defined similarly.
\end{definition}
Next, we introduce the notion of a decrescent function.
\begin{definition}\em{(\cite{SL})
A scalar function $V(x,\zeta)$  is said to be \textit{decrescent} if $V(0,\zeta)=0$, and if there exists a distance-invariant positive definite function $W(x)$ such that
\begin{align}
\forall\zeta\geq0,\quad V(x,\zeta)\leq W(x).
\end{align}
In other words, $V(x,\zeta)$ is decrescent if it is dominated by a distance-invariant positive definite function.}
\end{definition}
With the above definitions introduced we now state the classical Lyapunov stability theorem for non-autonomous systems.
\begin{theorem}\label{stability_theorem}\em{(\cite{SL})
If, in a ball $B_r(0)$ around the equilibrium point $0$, there exists a scalar function $V(x,\zeta)$ with continuous partial derivatives such that
\begin{enumerate}
\item[(i)] V is positive definite,
\item[(ii)] $V_{\zeta}$ is negative semi-definite.
\end{enumerate}
Then the equilibrium point $0$ is stable in the sense of Lyapunov.  If, furthermore,
\begin{enumerate}
\item[(iii)] $V$ is decrescent,
\end{enumerate}
then the origin is uniformly stable.}
\end{theorem}

With the stability of the linear system established, we turn our attention to the nonlinear system \eqref{system1}.  We seek to prove the stability of the origin for \eqref{system1} in the case where $n$ is even and ${\bar{y}_2}^{T}=\left(-\frac{1}{{\Omega}^{1/n}},0\right)$, leading to the system \eqref{system1} with the minus sign $(-)$.

Defining the Lyapunov function\begin{equation}\label{Lya}
   V(x)=-\frac{2\Omega}{(n+1)^2}\left[\left(x_1-\frac{1}{\Omega^{1/n}}\right)^{n+1}+\frac{1}{\Omega^{1+1/n}}\right]+\frac{2}{n+1}x_1+{x_2}^2,
   \end{equation} we then calculate the  derivative of $V$
\begin{align}\label{Vdot}
V_{\zeta}&=-\frac{2\Omega x_2}{n+1}\left(x_1-\frac{1}{\Omega^{1/n}}\right)^{n}+\frac{2x_2}{n+1}+\frac{2x_2}{n+1}\left[\Omega\left(x_1-\frac{1}{\Omega^{1/n}}\right)^n-1\right]-\frac{4{x_2}^2}{\zeta}\\ \nonumber
&=-\frac{4{x_2}^2}{\zeta}\leq 0.
\end{align}
  If $x_1<\frac{1}{\Omega^{1/n}}$,
then $-x_1\Omega^{1/n}\geq-1$.  This allows us to utilize the Bernoulli inequality $(1+u)^s\geq 1+su$, where $u\ge-1$, $s\in\mathbb{R}^{+}\setminus(0,1)$.
\begin{theorem}\label{Lyp_stable}\em{
The origin is a uniformly Lyapunov stable equilibrium point of system \eqref{system1} for $n$ even.}
\end{theorem}

\proof
We start by establishing that $V(x)$ is a locally positive definite function for $x\in B_{r}(0)$ where $r=\frac{2}{\Omega^{1/n}}$, and $\zeta\geq\zeta_0$.  We first consider the case where $x_1\leq\frac{1}{\Omega^{1/n}}$, and we apply the Bernoulli inequality above to obtain
\begin{align*}
V(x)&=\frac{2\Omega}{(n+1)^2}\frac{1}{\Omega^{1+1/n}}\left(1-x_1\Omega^{1/n}\right)^{n+1}-\frac{2}{(n+1)^2\Omega^{1/n}}+\frac{2}{n+1}x_1+{x_2}^2\\&\geq\frac{2}{(n+1)^2}\frac{1}{\Omega^{1/n}}\left(1-(n+1)x_1\Omega^{1/n}\right)-\frac{2}{(n+1)^2\Omega^{1/n}}+\frac{2}{n+1}x_1+{x_2}^2\\&=-\frac{2}{n+1}x_1+\frac{2}{n+1}x_1+{x_2}^2={x_2}^2\geq0.
\end{align*}
To consider the case where $\frac{1}{\Omega^{1/n}}\leq x_1\leq\frac{2}{\Omega^{1/n}}$  we note that the function
\begin{align*}
W(u)=-\frac{2\Omega}{(n+1)^2}\left[\left(u-\frac{1}{\Omega^{1/n}}\right)^{n+1}+\frac{1}{\Omega^{1+1/n}}\right]+\frac{2}{n+1}u
\end{align*}
satisfies
\begin{align*}
W_u(u)=-\frac{2\Omega}{n+1}\left(u-\frac{1}{\Omega^{1/n}}\right)^n+\frac{2}{n+1}.
\end{align*}
We then see that $W_u(u)\geq0$ for $x_1\in\left[\frac{1}{\Omega^{1/n}},\frac{2}{\Omega^{1/n}}\right]$, and since $W\left(\frac{1}{\Omega^{1/n}}\right)=\frac{2}{n+1}\left(1-\frac{1}{n+1}\right)>0$, we conclude that $W(u)\geq0$ on $\left[\frac{1}{\Omega^{1/n}},\frac{2}{\Omega^{1/n}}\right]$ which in turn implies that $\left.V(x)\right|_{B_r(0)}\geq0$.

Since $V$ is distance-invariant it follows that $V$ is a locally positive definite function, moreover it is evident that $V$ is also a decrescent function.  Therefore using Theorem~\ref{stability_theorem} together with $V(0)=0$, and $V_{\zeta}\leq0$ we conclude that the origin is uniformly Lyapunov stable. \eproof
\begin{theorem}\label{asympt}\em{The equilibrium point $0$ of system \eqref{system1} is asymptotically stable when $n$ is even.
}
\end{theorem}
\proof
 To study the asymptotic stability of the origin, we note that on the domain $D=\left\{x_1\leq\frac{2}{\Omega^{1/n}}\right\}$ we have that $\left.V\right|_{D}$ is positive definite and $\left.V_{\zeta}\right|_{D}\leq0$.  Unlike in the autonomous case where LaSalle's invariance theorem can be used to find a maximally invariant set that the trajectories converge to, in the nonautonomous case it is less clear how to define such a set.  In fact since $-V_{\zeta}$ is not a locally positive definite function, Theorem~8.4 in \cite{HK} does not directly apply.  We can still use the ideas from Theorem~8.4 to construct an invariant set in $D$. If we consider the closed ball of radius $r=\frac{2}{\Omega^{1/n}}$ centered at the origin, then $B_{r}(0)\subset D$.  We introduce the following distance invariant set $\mathcal{B}_{\alpha}=\left\{x\in B_{r}(0)\mid V(x)\leq \alpha \right\}$, where $\alpha<\min_{\left\|x\right\|=r}V(x)$.  To calculate $\min_{\left\|x\right\|=r}V(x)$ we proceed as follows: on $\overline{B_{r}(0)}$ we can write $V(x_1)= -\frac{2\Omega}{(n+1)^2}\left[\left(x_1-\frac{1}{\Omega^{1/n}}\right)^{n+1}+\frac{1}{\Omega^{1+1/n}}\right]+\frac{2}{n+1}x_1+r^2-{x_1}^2$ with the derivative  $V_{x_1}=-\frac{2}{n+1}\left(1-\Omega^{1/n}x_1\right)^n+\frac{2}{n+1}-2x_1$.  We then see that $V_{x_1}(0)=0$, and note that $V_{x_1x_1}=\frac{2\Omega^{1/n}}{1+1/n}\left(1-\Omega^{1/n}x_1\right)^{n-1}-2$.  If $-\frac{2}{\Omega^{1/n}}\leq x_1\leq \frac{1}{\Omega^{1/n}}\left[1-\left(\frac{1+1/n}{\Omega^{1/n}}\right)^{\frac{1}{n-1}}\right]$ then $V_{x_1x_1}\geq 0$, and  for $x_1\geq \frac{1}{\Omega^{1/n}}\left[1-\left(\frac{1+1/n}{\Omega^{1/n}}\right)^{\frac{1}{n-1}}\right]$ then $V_{x_1x_1}\leq 0$. Thus, $x_1=0$ is the only critical point. Since $V(\frac{2}{\Omega^{1/n}},0)$ is less than both $V(0,0)$ and $V(-\frac{2}{\Omega^{1/n}},0)$ then $\min_{\left\|x\right\|=r}V(x)=V(\frac{2}{\Omega^{1/n}},0)$.

Returning to the set $\mathcal{B}_{\alpha}$, we have that  $\mathcal{B}_{\alpha}\subset B_{r}(0)\subset D$ for all $\zeta\geq\zeta_0$ which implies that $\mathcal{B}_{\alpha}$ is bounded.  Since $V_{\zeta}\leq0$ in $D$ we have that for $\zeta_0>0$ the solution starting at $(x_0,\zeta_0)$ stays in $\mathcal{B}_{\alpha}$ for all $\zeta\geq\zeta_0$, hence the solution is bounded for all $\zeta\geq\zeta_0$.  Moreover, from \eqref{Vdot} we have
\begin{align}\label{asym1}
V(x(\zeta))-V(x(\zeta_0))=-4\int_{\zeta_0}^{\zeta}\frac{{x_2}^2(s)}{s}\;ds.
\end{align}
Since $V$ is bounded below on $D$, and $V_{\zeta}\leq 0$ we have that $\lim_{\zeta\rightarrow\infty}V(x(\zeta),\zeta)=V_{\infty}$ exists,  and $V_{\infty}\leq V(x(\zeta_0),\zeta_0)$.  Taking  $\zeta\rightarrow\infty$ in \eqref{asym1} yields
\begin{align}
V(x(\zeta_0))-V_{\infty}=4\lim_{\zeta\rightarrow\infty}\int_{\zeta_0}^{\zeta}\frac{{x_2}^2(s)}{s}\;ds.
\end{align}
Next, we claim that $\lim_{\zeta\rightarrow\infty}{x_2}^2(\zeta)=0$. By  contradiction,  suppose that $\lim_{\zeta\rightarrow\infty}{x_2}^2(\zeta)=L$ where $L\neq0$  (note that $\lim_{\zeta\rightarrow\infty}x_2(\zeta)$ exists since $V_{\infty}$ exists). Then for $\epsilon=\frac{L}{2}$ there exists $\zeta_1>0$ such that ${x_2}^2(\zeta)>L/2$ for $\zeta\geq\zeta_1$.  This implies that
\begin{align}\label{asym2}
\lim_{\zeta\rightarrow\infty}\int_{\zeta_1}^{\zeta}\frac{{x_2}^2(s)}{s}\;ds\geq\frac{L}{2}\lim_{\zeta\rightarrow\infty}\int_{\zeta_1}^{\zeta}\frac{1}{s}\;ds=\infty.
\end{align}
Eq. \eqref{asym2} implies that $V(x(\zeta_0))-V_{\infty}$ is divergent which is a contradiction, therefore $\lim_{\zeta\rightarrow\infty}x_2(\zeta)=0$.  Moreover, ${x_2}_{\zeta}$ is uniformly continuous since
\begin{align*}
{x_2}_{\zeta\zeta} =\frac{\Omega}{1+1/n}\left(x_1-\frac{1}{\Omega^{1/n}}\right)^{n-1}x_2-\frac{2}{\zeta(n+1)}\left[\Omega\left(x_1-\frac{1}{\Omega^{1/n}}\right)^{n}-1\right]+\frac{6x_2}{\zeta^2}
\end{align*}
is bounded because $x_1$ and $x_2$ are bounded functions.  Thus, from Barbalat's lemma \cite{SL} we see that $\lim_{\zeta\rightarrow\infty}{x_2}_{\zeta}=0$ which implies from \eqref{system1} that \\ $\lim_{\zeta\rightarrow\infty}\Omega\left(x_1-\frac{1}{\Omega^{1/n}}\right)^n-1=0$, and hence $\lim_{\zeta\rightarrow\infty}x_1(\zeta)=0$.
\eproof

At this point, we have established that the origin is asymptotically stable for \eqref{system1}.  With this in mind, it is of interest to estimate the basin of attraction for the origin by using the ideas from the proof of Theorem~\ref{asympt}.  We start with noting that the constant $\alpha$ from the above proof satisfies $\alpha<V\left(\frac{2}{\Omega^{1/n}},0\right)=\frac{4n}{\Omega^{1/n}(n+1)^2}$.
Setting $\alpha=\frac{4n}{\Omega^{1/n}(n+1)^2}-\delta$ for $\delta$ sufficiently small, the set $\mathcal{B}_{\alpha}$ can be expressed as $\mathcal{B}_{\delta}=\left\{x\in \left.B_{\frac{2}{\Omega^{1/n}}}(0)\right| V(x)\leq \frac{4n}{\Omega^{1/n}(n+1)^2}-\delta\right\}$.  Therefore, for an initial condition of the form $x_0^T=\left(\frac{1}{\Omega^{1/n}}-\epsilon,0\right)$ for $\epsilon>0$ arbitrarily small, we can find $\delta>0$ such that $\mathcal{B}_{\delta}$ is a positively invariant set containing $x_0$ and hence the flow to \eqref{system1} with initial condition $x_0$ converges asymptotically to the origin.

To continue our stability analysis, we consider the equilibrium point ${\bar{y}_{0}}^{T}=\left(\frac{1}{{\Omega}^{1/n}},0\right)$,
which is present as the only equilibrium when $n$ is odd, and is one of two equilibria for $n$ even.  It is our intention to establish $\bar{y}_{0}$ as unstable.  To accomplish this we apply one of the instability theorems found in \cite{SL}.  The notion of instability used here is taken from \cite{WH}, that is the equilibrium of the system \eqref{system1} is called unstable if it is not stable.  To this end, there exists an $\epsilon>0$ arbitrarily small such that no $\delta>0$ exists ensuring Lyapunov stability.  This means there exists a $x_{0}$ arbitrarily close to 0 and a sequence $\zeta_n\rightarrow\infty$ so that $\left|x(\zeta_n)\right|\geq\epsilon$ for all $\zeta_n\geq\zeta_0$,
where $x(\zeta_0)=x_0$.
\begin{theorem}\label{unstable}\em{\cite{SL} If, in a certain neighborhood $\mathcal{B}_0$ of the origin, there exists a continuously differentiable, decrescent scalar function $V(x,\zeta)$ such that
\begin{enumerate}\label{theorem_inst}
\item[(i)] $V(0,\zeta)=0,\quad \zeta\geq\zeta_0$,
\item[(ii)]$V(x,\zeta_0)$ can assume strictly positive values arbitrarily close to the origin,
\item[(iii)]$V_\zeta(x,\zeta)$ is positive definite (locally in $\mathcal{B}_0$),
\end{enumerate}
then the equilibrium point $0$ at distance $\zeta_0$ is unstable.}
\end{theorem}
Recall for $n$ odd, the only real equilibrium point is ${\bar{y}_{0}}^{T}=\left(\frac{1}{{\Omega}^{1/n}},0\right)$, and when $n$ is even the real equilibrium point lying on the negative horizontal axis is ${\bar{y}_{1}}^{T}=\left(-\frac{1}{{\Omega}^{1/n}},0\right)$. For  the odd case the system obtained by shifting $x=y-\bar{y}_{0}$ is \eqref{system1} with the $(+)$ sign.  In what follows we show that the origin is unstable in the case where $n$ is odd.
\begin{theorem}\label{unstable2}\em{
The equilibrium point $0$ of \eqref{system1} is unstable for odd $n$.}
\end{theorem}
\proof We apply the above instability Theorem \ref{unstable}.  We define the following scalar valued Lyapunov function
\begin{equation}V(x,\zeta)=\left(\Omega\left(x_1+\frac{1}{\Omega^{1/n}}\right)^n-1\right)x_2+\frac{(n+1){x_2}^2}{\zeta},
\end{equation}
noting that $V(0,\zeta)=0$, and also that $V(x,\zeta)$ is a decrescent function. Furthermore, for $x_1=0$ and $x_2\neq 0$, $V(x,\zeta)$ is strictly positive, hence condition (ii) in Theorem~\ref{theorem_inst} is satisfied. For the  derivative we obtain
\begin{align*}
V_{\zeta}&=\frac{1}{n+1}\left(\Omega\left(x_1+\frac{1}{\Omega^{1/n}}\right)^n-1\right)^2+n\Omega {x_2}^2\left(x_1+\frac{1}{\Omega^{1/n}}\right)^{n-1}-\frac{5(n+1){x_2}^2}{\zeta^2}\\&=\frac{1}{n+1}\left(\Omega\left(x_1+\frac{1}{\Omega^{1/n}}\right)^n-1\right)^2+{x_2}^2\left(n\Omega\left(x_1+\frac{1}{\Omega^{1/n}}\right)^{n-1}-\frac{5(n+1)}{\zeta^2}\right).
\end{align*}
Since $\left(x_1+\frac{1}{\Omega^{1/n}}\right)^{n-1}\geq\left(-\frac{1}{2\Omega^{1/n}}+\frac{1}{\Omega^{1/n}}\right)^{n-1}=\frac{1}{2^{n-1}\Omega^{1-1/n}}$, then by taking \\ ${\zeta_0}^2=1+\frac{5(1+1/n)2^{n-1}\Omega^{1-1/n}}{\Omega}$
we have that
\begin{align}\nonumber
V_{\zeta}&\geq \frac{1}{n+1}\left(\Omega\left(x_1+\frac{1}{\Omega^{1/n}}\right)^n-1\right)^2+{x_2}^2\left(n\Omega\left(x_1+\frac{1}{\Omega^{1/n}}\right)^{n-1}-\frac{5(n+1)}{\zeta_0^2}\right)\\&
=V_0(x)\geq 0,
\end{align}
provided $\left\|x\right\|<r=\frac{2}{\Omega^{1/n}}$.  So $V_0(x)$ is a distance-invariant locally positive definite function, therefore (iii) is satisfied, and Theorem~\ref{unstable} implies that $0$ is unstable at $\zeta_0$.  Thus, we can conclude that $0$ is unstable for all  $\zeta\geq\zeta_0$. \eproof
\section{Orbital modes}
In this section, we discuss the effects of $n$ and $\Omega$ on the solutions of (\ref{eq1}). The formation of stable orbital modes in the density function was observed in  rotary clouds of  atoms \cite{Wal,Sesko,Bag,Arn}, and was previously believed to be a consequence of rotation in the cloud. 
The analysis of the previous section reveals bounded solutions when  $n$ is  even which are periodic as long as the starting initial conditions are to the left of the unstable positive fixed point ${\bar{y}_2}^T$, and are inside the basin of attraction $\mathcal{B}_\alpha$. Since we use  the initial condition $\theta(\zeta_0)=1$, and  to be inside of the basin we require that  $1<\frac{1}{\Omega^{1/n}}$. This  yields to $0<\Omega<1$, which was also claimed numerically, and by linear analysis  by \cite{Terc}. The solutions are unbounded when $n$ is odd since there is only one unstable equilibrium given by ${\bar{y}_0}^T$, or if the initial conditions are outside of the basin of attraction $\mathcal{B}_\alpha$. In \cite{Rod1} and \cite{Terc}, the authors discuss orbital modes for the cases $\gamma=1,2,4$.   However, the case $\gamma=1$ is discussed in Section~1, and $\gamma=2$ follows from the conclusion of Theorem \ref{unstable2}.  It is relevant to note that the case $\gamma=\frac{3}{2}$ is studied in \cite{Rod2} of which the stability follows from Theorem \ref{asympt}.

\begin{figure}[ht!]
     \begin{center}
\includegraphics[width=\textwidth]{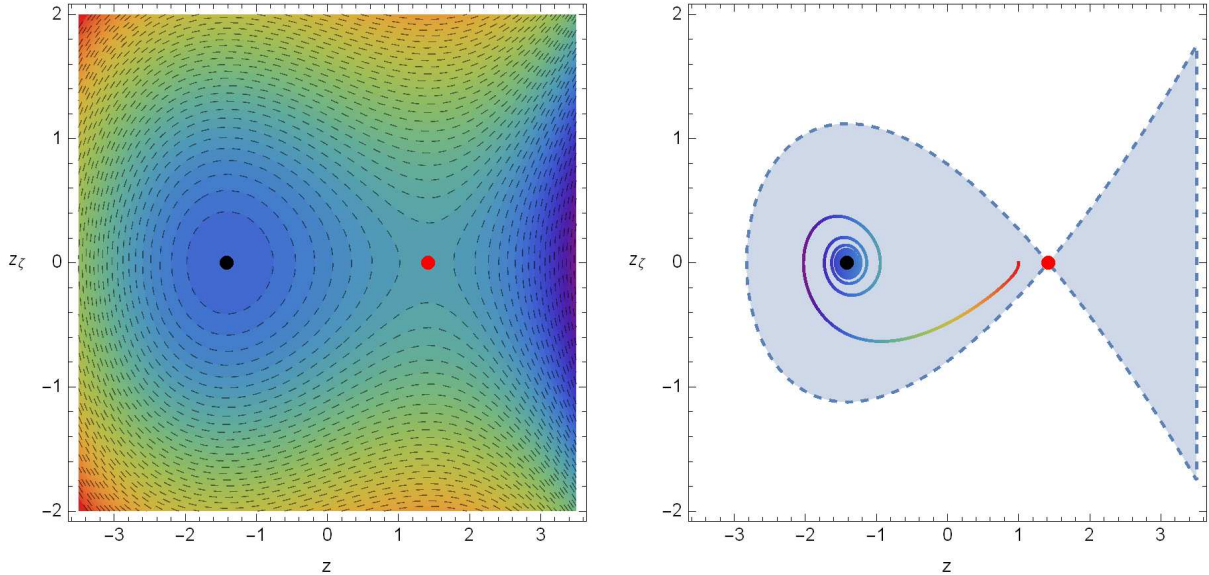}
\caption{Left panel: translations of the level sets of the Lyapunov function $V(x)$ given by (\ref{Lya}). Right panel: basin of attraction for periodic solutions of (\ref{system1}). The fixed points are indicated by the black and red  dots, left fixed point is always stable while the right fixed point is always unstable. Simulations must start inside of the basin of attraction which require that $0<\Omega<1$. Also, the necessary conditions to have two fixed points is that $n$ is even. A typical trajectory starts close and to the left of the unstable fixed point and converges to the stable fixed point. }
\label{fig1}
\end{center}
\end{figure}

The left panel of Fig. \ref{fig1} depicts a sublevel set for the  function $V\left(x+\bar{y}_{1}\right)$, which are the translations of the level sets of the Lyapunov function $V(x)$.  The  right panel corresponds to an asymptotically stable solution to (\ref{eq5}) with $n=2$, $\Omega=0.5$, and left stable fixed point ${\bar{y}_1}^T=\left(-\frac{1}{\Omega^{1/n}},0\right)$. We start from initial condition  $x_0^T=\left(\frac{1}{\Omega^{1/n}}-\epsilon,0\right)$ located  to the left of the right unstable fixed point ${\bar{y}_2}^T=\left(\frac{1}{\Omega^{1/n}},0\right)$.  The shaded  region in the right panel represents the sublevel set $\left\{x\in \mathbb{R}^2\mid V(x+\bar{y}_{1})\leq \alpha\right\}$, we take the bounded component of this set, then translate it to the right by $\bar{y}_1$ to obtain $\mathcal{B}_{\alpha}$ which is used to estimate the basin of attraction for the origin of system (\ref{system1}).

\begin{figure}[ht!]
     \begin{center}
\includegraphics[width=\textwidth]{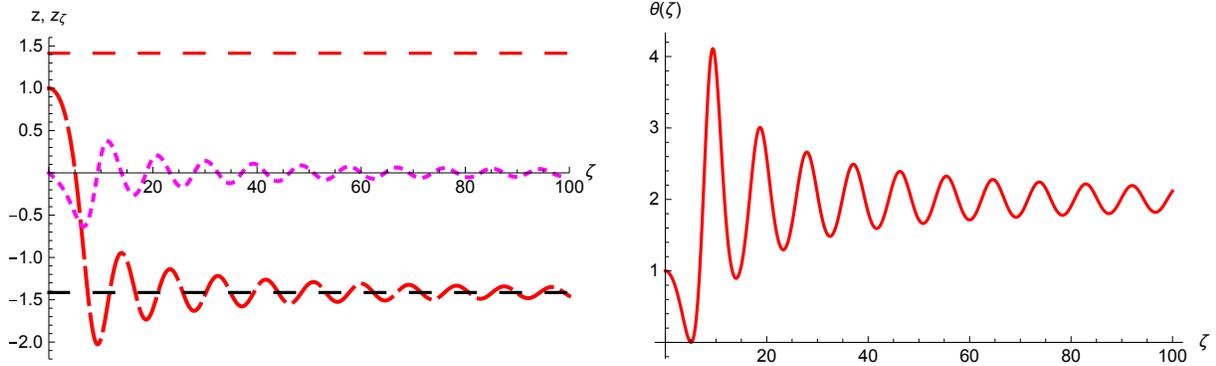}
\caption{Left panel: solutions  $z(\zeta)$, $z_{\zeta}(\zeta)$ of (\ref{eq5}),  with the red dashed line representing the right (unstable) fixed point, and the black  dashed line the left (stable) fixed  point. The red dashed function $z(\zeta)$ and the magenta dotted function $z_\zeta(\zeta)$ represent the trajectory from the right panel of Fig. \ref{fig1}.   Right panel: density function  $\theta(\zeta)=z^2$  of  (\ref{eq1}), with $\gamma=\frac{3}{2}, \Omega=0.5$ of the solution $z(\zeta)$ from the left panel.}
\label{fig2}
\end{center}
\end{figure}

\begin{figure}[ht!]
     \begin{center}
\includegraphics[width=0.6\textwidth]{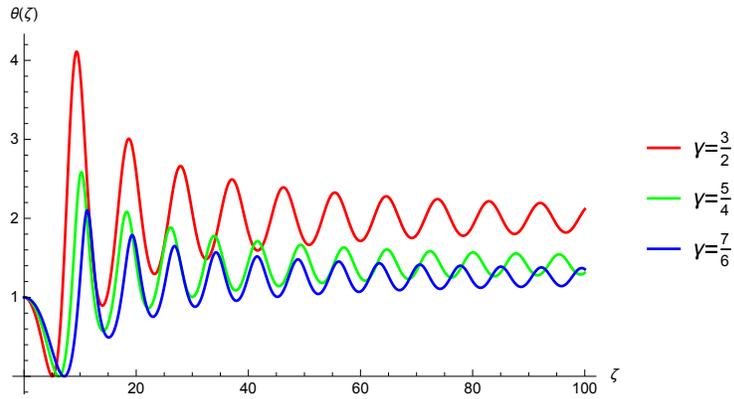}
\caption{Stable periodic solutions  $\theta(\zeta)$  of  (\ref{eq1})  with $\Omega=0.5$, and three values of the  parameter $\gamma=\frac{3}{2},\frac{5}{4}$, and $ \frac{7}{6}$. The top curve is  the solution from the right panel of  Fig. \ref{fig2}.}
\label{fig3}
\end{center}
\end{figure}

The left panel of Fig. \ref{fig2} shows the solutions  $z(\zeta)$, $z_{\zeta}(\zeta)$ of (\ref{eq5}) with the red dashed line representing the right (unstable) equilibrium, and the black  dashed line the left (stable) equilibrium point.  The right figure depicts the density function  $\theta(\zeta)$  of  (\ref{eq1}), with $\gamma=\frac{3}{2}, \Omega=0.5$, and initial conditions $\theta(\zeta_0)=1$, $\theta_{\zeta}(\zeta_0)=0$. Since $\zeta=0$ is a singular point of (\ref{eq1}), for numerical simulations, we used $\zeta_0=0.001$.
When $n=2,4,6$ which gives $\gamma=\frac 32, \frac 54, \frac 76$, and starting with the same  initial conditions,  we obtain periodic solutions that are depicted in Fig. \ref{fig3}. The boundary $\zeta^*$ is given by $\theta(\zeta)=0$,  is increasing as $\gamma$ is increasing, and represents the nondimensional  radius for which the density of atoms is zero.

\begin{figure}[ht!]
     \begin{center}
\includegraphics[width=0.6\textwidth]{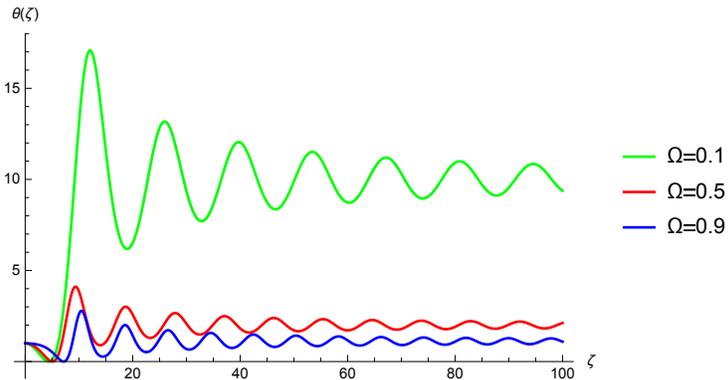}
\caption{Stable periodic solutions $\theta(\zeta)$  of  (\ref{eq1})  with $\gamma=\frac{3}{2}$,  and  three values of the  parameter $\Omega=0.1,0.5$, and $0.9$.  The middle curve is  the solution from the right panel of  Fig. \ref{fig2}.}
\label{fig4}
\end{center}
\end{figure}

For a fixed adiabatic index, and with the number of atoms that increases  then $\Omega$ increases, thus the central core density increases. This  was observed experimentally when a single ring turns into a ring with a central core, causing the cloud to rotate faster \cite{Bag}, as we can see in Fig. \ref{fig4}.
\section{Conclusion}
A detailed stability analysis of a generalized Lane-Emden equation recently elaborated in the physics of trapped atomic clouds, and previously in astrophysics has been presented.  The linearized form  of the corresponding nonautonomous first order system has been shown to be locally asymptotically stable, and the Lyapunov indirect approach has been used to construct a Lyapunov  function for the nonlinear system.  When $n$ is even there are two equilibrium points $\bar{y}_{1}<\bar{y}_{2}$, and we showed that the left equilibrium point $\bar{y}_1$ is asymptotically stable. Using the Lyapunov function we provide an estimate for the basin of attraction in which initial conditions should be picked  that yield periodic solutions.  For the other equilibrium point $\bar{y}_{2}$, we provide an instability result showing that in the case where $n$ is odd this equilibrium is unstable, and when $n$ is even the same equilibrium $\bar{y}_{0}=\bar{y}_{2}$  is also unstable.

This  analysis  yields  a special set of solutions for the density of atoms of the generalized Lane-Emden equation obtained when $n$ is even, and initial conditions are inside a region of attraction to the left of the stable equilibrium. These periodic solutions (orbital modes) have been observed experimentally previously in rotating ultracold atomic clouds, and are demonstrated analytically to be stable.

\end{document}